\documentclass[11pt]{article} % use larger type; default would be 10pt

\usepackage[utf8]{inputenc} % set input encoding (not needed with XeLaTeX)
\usepackage{amsfonts}
\usepackage{booktabs}
\usepackage{siunitx}

\usepackage{tikz}
\usetikzlibrary{patterns,shapes,decorations}

\usepackage[margin=0.8in]{geometry}% to change the page dimensions
\usepackage{setspace}
\usepackage{kotex}
\usepackage{graphicx,color} % support the \includegraphics command and options
\usepackage{amsmath,amssymb}
\usepackage{amsthm}
\usepackage{hyperref} 
\usepackage[export]{adjustbox}
\usepackage{footnote,xcolor}
\hypersetup{linkbordercolor=green}
\usepackage{booktabs} % for much better looking tables
\usepackage{array} % for better arrays (eg matrices) in maths
\usepackage{paralist} % very flexible & customisable lists (eg. enumerate/itemize, etc.)
\usepackage{verbatim} % adds environment for commenting out blocks of text & for better verbatim
\usepackage{subfig} % make it possible to include more than one captioned figure/table in a single float
\usepackage{fancyhdr} % This should be set AFTER setting up the page geometry
\usepackage{sectsty}
\allsectionsfont{\sffamily\mdseries\upshape} % (See the fntguide.pdf for font help)
\usepackage[round]{natbib}   % omit 'round' option if you prefer square brackets

\usepackage[titles,subfigure]{tocloft} % Alter the style of the Table of Contents

\newcommand{\E}{\mathbf{E}}

\newcommand{\indep}{\perp \!\!\! \perp}
\newcommand{\for}{\text{for }}
\newtheorem{assumption}{Assumption}

\newtheorem{lemma}{Lemma}
\newtheorem{proposition}{Proposition}

 \usepackage{titlesec}
 
\usepackage{mdframed}
\usetikzlibrary{shapes,decorations,arrows,calc,arrows.meta,fit,positioning}
\tikzset{
    -Latex,auto,node distance =1 cm and 1 cm,semithick,
    state/.style ={ellipse, draw, minimum width = 0.7 cm},
    point/.style = {circle, draw, inner sep=0.04cm,fill,node contents={}},
    bidirected/.style={Latex-Latex,dashed},
    el/.style = {inner sep=2pt, align=left, sloped}
}

\titleformat*{\section}{\Large\bfseries}
\titleformat*{\subsection}{\large\bfseries}
\titleformat*{\subsubsection}{\large\bfseries}
\titleformat*{\paragraph}{\small\bfseries}
\titleformat*{\subparagraph}{\small}
\usepackage{enumitem}

\usepackage{siunitx}
\usetikzlibrary{positioning}
\tikzset{mynode/.style={draw,text width=1in,align=center}
}

\newlist{thmlist}{enumerate}{1}
\setlist[thmlist]{label=(\roman{thmlisti}),noitemsep}

\usepackage{thmtools}

\usepackage{hyperref}
\hypersetup{colorlinks=true, linkcolor=blue, filecolor=blue, urlcolor=blue, citecolor=blue} % use for hypertext
\usepackage[colorinlistoftodos]{todonotes}

         \renewenvironment{abstract}
 {\small
  \begin{center}
  \bfseries \abstractname\vspace{-.5em}\vspace{0pt}
  \end{center}
  \list{}{%
    \setlength{\leftmargin}{5mm}% <---------- CHANGE HERE
    \setlength{\rightmargin}{\leftmargin}%
  }%
  \item\relax}
 {\endlist}
\begin{document}
%\title{Treatment effect analysis under social interactions}
%\maketitle
%\tableofcontents
%=============================================================================%%=============================================================================%
\title{On the Use of Instrumental Variables in Mediation Analysis}
\author{Bora Kim}
\maketitle
%\tableofcontents

\begin{abstract}
 Empirical researchers are often interested in not only whether a treatment affects an outcome of interest, but also how the treatment effect arises. Causal mediation analysis provides a formal framework to identify causal mechanisms through which a treatment affects an outcome. The most popular identification strategy relies on so-called sequential ignorability (SI) assumption which requires that there is no unobserved confounder that lies in the causal paths between the treatment and the outcome. Despite its popularity, such assumption is deemed to be too strong in many settings as it excludes the existence of unobserved confounders. This limitation has inspired recent literature to consider an alternative identification strategy based on an instrumental variable (IV). This paper discusses the identification of causal mediation effects in a setting with a binary treatment and a binary instrumental variable that is both assumed to be random. We show that while IV methods allow for the possible existence of unobserved confounders, additional monotonicity assumptions are required unless the strong constant effect is assumed. Furthermore, even when such monotonicity assumptions are satisfied, IV estimands are not necessarily equivalent to target parameters. 
\end{abstract}
\pagenumbering{arabic}

 %=============================================================================%
 % how such a causal relationship arises (imai)
 \newpage
\section{Introduction}
Understanding causal mechanisms through which a treatment or an intervention ($D$) affects an outcome ($Y$) is a fundamental goal of social science. Researchers are interested not only in identifying whether there is a treatment effect, but also in understanding how such a treatment effect arises. Suppose, for example, that an early childhood program ($D$) shows a positive effect on an adult outcome ($Y$).  An important question that follows is whether and to what extent such effect can be attributed to a change in an educational achievement that is itself induced by the program,
 %or whether there is another pathway that is not operating through the educational achievement 
(see \cite{early}).  Causal mediation analysis offers a formal framework to uncover causal mechanism, a set of casual pathways connecting $D$ and $Y$, underlying observed treatment effects. Specifically, it aims at decomposing a total effect of $D$ on $Y$ into an indirect effect operating through a third variable called “mediator”, $M$, (e.g., through years of education) and a direct effect that does not operate through that mediator (e.g., through personality traits). Understanding the mechanism of causal effects allows one to design more effective policies which may involve altering specific causal pathways.

Decomposing the total treatment effect into direct and indirect effects is a challenging task. Even with the “gold standard” randomized controlled trial (RCT) where the treatment is randomized, direct and indirect effects are not identified without further assumptions since the mediator, a post-intervention outcome, is in general non-random, thus making it difficult to identify the causal effect of the mediator on the outcome. The so-called black box critique of RCTs illustrates the difficulty of performing causal mediation analysis. 

The additional assumption that is commonly invoked in the literature in order to identify direct and indirect effects is so-called ``sequential ignorability" (SI) assumption, which is essentially a selection-on-observables assumption on both $D$ and $M$. Under the sequential ignorability assumption, $D$ and $M$ can be considered as-if random after controlling for the relevant set of observable covariates. To illustrate the identification power of SI, let us consider a linear regression model for the random $D$ as follows: for simplicity, let us assume that there is no covariate:
 \begin{eqnarray} \label{LSEM1}
 Y_i=b_0+b_1D_i+b_2M_i+b_3D_iM_i+u_i,\quad 
 M_i=a_0+a_1D_i+v_i.
 \end{eqnarray}
 Here, $(u_i,v_i)$ are unobservables. SI assumes that $corr(u_i,D_i)=corr(u_i,M_i)=0$ and $cov(D_i,v_i)=0$, which in turn implies $corr(u_i,v_i)=0$ after controlling for relevant covariates $X_i$. Under these assumptions, coefficients $(a,b)$ can be consistently estimated using least-squares method and direct and indirect effects are then estimated as a function of estimated coefficients. 
 
Sequential ignorability assumption is arguably strong as it excludes unobserved confounders affecting both $M$ and $Y$, which is unlikely to hold in many realistic settings. For instance in early childhood programs, SI fails if there is an unobserved individual trait such as perseverance which affects both education levels ($M$) and earnings ($Y$) regardless of the program participation ($D$). In the context of linear regression model above, even after controlling a rich set of covariates, we may still have $corr(u_i,v_i)\not=0$, leading to failure of SI.

 Recently, several papers have proposed an alternative identification strategy based on instrumental variables (IV) to address the possible existence of unobserved confounders. Theoretical papers include \cite{frolich}, \cite{dippel}, \cite{imai2013} and \cite{mealli}. Empirical paper include \cite{chen} and \cite{dippel2}. 
 Assuming that $D$ random, the method supposes the existence of valid IV, denoted by $Z$, for $M$ in the sense that (i) $Z$ is exogenous to both $D$ and $M$, and (ii) $Z$ affects $Y$ only through $M$. The method attempts to exploit the resulting exogenous variation in $M$ generated by $Z$. In a linear regression model, we now have
   \begin{eqnarray} \label{LSEM2}
 Y_i=b_0+b_1D_i+b_2M_i+b_3D_iM_i+u_i,\quad 
 M_i=a_0+a_1D_i+a_2Z_i+a_3 D_iZ_i+v_i
 \end{eqnarray}
where coefficients are estimated using IV methods where $(D_i,M_i,D_iM_i)$ is instrumented by $(D_i,Z_i,D_iZ_i)$. This produces a consistent estimator of the coefficients $(a,b)$, and thus (in)direct effects as well even when $corr(u_i,v_i)\not=0$.

While IV methods can address unobserved confounders in the relationship between $M$ and $Y$, they generally involve additional assumptions other than exogeneity and exclusion restriction once we attempt to move beyond the linear regression model. In a non-mediation setting where we aim to identify the causal effect of $D$ on $Y$ using $Z$ as an IV for $D$, \cite{IA} shows that either (i) constant effect assumption or (ii) so-called monotoncity assumption is required for IV estimand to have a causal interpretation. \cite{IA} also shows that under the heterogeneous effect setting, even when the monotonicity assumption is satisfied, the average treatment effect is not identified; instead we can only identify the average treatment effect for a certain subpopulation known as the compliers. 

Similar arguments are expected to hold in mediation settings as well, albeit more complex. However, despite its increasing popularity, there is no formal result outlining the formal identification result under IV in a mediation setting. This paper fills this gap by formally deriving a set of required assumptions  in the context of randomized treatment.  Our first result shows that when the constant effect is assumed, IV estimator for (in)direct effects identifies the true (in)direct effects. Such constant effect amounts to assuming that the linear model above is correctly specified: that the coefficients $(a,b)$ are truly constant after controlling for all observable covariates. 

When such assumption is violated due to random (unobserved) coefficients (that is, the true model has $(a_i,b_i)$ rather than $(a,b)$), we show that certain monotonicity assumptions are required for IV estimands to have causal interpretations, in the sense that it identifies a positively weighted averages of some subgroup effects. Specifically, we need $M$ to be partially monotonic in both $(D,Z)$ similar to \cite{IA}'s no-defier assumption. 
Finally, we show that even when such partial monotonicity assumption is satisfied, there is no guarantee that IV estimands are informative on the target parameter, similar to \cite{IA}'s result that IV methods identify LATE, not ATE.

%---------------------------------------------------------------------------------------------------------------------------------------%

%=============================================================================%
\section{Framework and Identifiability}
%=============================================================================%
%=============================================================================%
\begin{figure}[!h]
\caption{Mediation Diagram}  \label{f1}
\centering
\begin{tikzpicture}
    \node[mynode] (m){Mediator ($M$)};
    \node[mynode,below left=of m](a) {Treatment ($D$)};
    \node[mynode,below right=of m](b) {Outcome ($Y$)};
    \draw[-latex] (a.north) -- node[auto,font=\footnotesize] {indirect} (m.west);
    \draw[-latex] (m.east) -- node[auto,font=\footnotesize] {effect} (b.north);
    \draw[-latex] (a.east) --
            node[below=3mm,font=\footnotesize,align=center] {direct effect}
                 (b.west);
  \end{tikzpicture}
\end{figure}
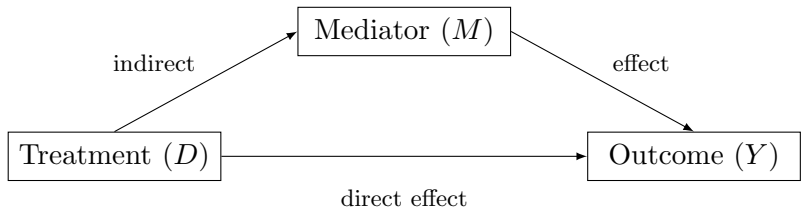
%---------------------------------------------------------------------------------------------------------------------------------------%
% diagram?
%=============================================================================%
The aim of causal mediation analysis is to quantify the extent to which the effect of a treatment on an outcome is mediated by a third variable, called ``mediator". Let us decompose a \emph{total effect} of a treatment on an outcome into an \emph{indirect effect} (or \emph{mediated effect}) which operates through the mediator, and a \emph{direct effect} (or \emph{unmediated effect}) which does not operate through the mediator, as depicted in Figure \ref{f1}.

Throughout the paper we consider a simple case of a binary treatment and a binary mediator. For each individual $i$, let $D_i\in\{0,1\}$ be an indicator of treatment (1: treated, 0: not treated), $M_i\in\{0,1\}$ be a binary mediator and $Y_i\in\mathbb{R}$ be an outcome of interest. We observe $(D_i,M_i,Y_i)$ for a random sample of individuals. For simplicity, let us suppress the individual index $i$. In addition, we abstract from any covariates and implicitly condition on them. 

%---------------------------------------------------------------------------------------------------------------------------------------%
\subsection{Potential outcomes and causal effects}
%---------------------------------------------------------------------------------------------------------------------------------------%
Following the literature, we define indirect and direct effects using the potential outcomes (or counterfactual) framework  (see, for example, \cite{pearl}).
Let $M_d$ denote the potential mediator value when the treatment is set to $D=d$. 
 Let $Y_{d,m}$ be the potential outcome when the treatment is set to $D=d$ and the value of mediator is set to $M=m$. Similarly, $Y_{d,M_{d'}}$ is the potential outcome where the treatment is set to $D=d$ while the mediator is set to $M=M_{d'}$, i.e., its potential value that would take under the treatment state $d'$. In this way, for each individual, we have two potential mediators, $(M_1,M_0)$, and four potential outcomes, $(Y_{1,M_1}, Y_{1,M_0}, Y_{0,M_1},Y_{0,M_0})$. Only one of each is observed. Realized outcome $Y$ and realized mediator $M$ satisfy the following conditions:
 $$M=M_D=DM_1+(1-D)M_0,\quad Y=Y_{D,M_D}=DY_{1,M_1}+(1-D)Y_{0,M_0}.$$ Note that  a counterfactual $Y_{d,M_{d'}}$ for $d\not=d'$ is \emph{never} observed in data, unless an individual has $M_{d'}=M_{d}$. Since only one of $(M_1,M_0)$ is revealed by data, it is not known a priori whether a certain individual has $M_1=M_0$ or not. 

In this paper we focus on mean effects. \emph{Average Total Effect} (ATE) of treatment is defined as follows:
\begin{eqnarray*}
ATE=\E[Y_{1,M_1}-Y_{0,M_0}]. 
\end{eqnarray*}
Note that when the treatment is randomized, $ATE$ is easily identified by $\E[Y|D=1]-\E[Y|D=0]$.

There are two ways to decompose $ATE$: First,
\begin{eqnarray*}
ATE=\underbrace{\E[Y_{\textcolor{red}{1},M_{1}}-Y_{\textcolor{red}{1},M_{0}}]}_{NIE_{\textcolor{red}{1}}}+\underbrace{\E[Y_{1,M_{\textcolor{red}{0}}}-Y_{0,M_{\textcolor{red}{0}}}] }_{NDE_{\textcolor{red}{0}}}.
\end{eqnarray*}
Second,
\begin{eqnarray*}
ATE=\underbrace{\E[Y_{1,M_{\textcolor{red}{1}}}-Y_{0,M_{\textcolor{red}{1}}}]}_{NDE_{\textcolor{red}{1}}}+\underbrace{\E[Y_{\textcolor{red}{0},M_{1}}-Y_{\textcolor{red}{0},M_{0}}]}_{NIE_{\textcolor{red}{0}}}.
\end{eqnarray*}
Following \cite{pearl}, we define \emph{natural} direct effects (NDE) as follows:
\begin{eqnarray}\label{DE}
NDE_{\textcolor{red}{d}} =^{def} \E[Y_{1,M_{\textcolor{red}{d}}}-Y_{0,M_{\textcolor{red}{d}}}],\quad\for d=0,1
\end{eqnarray} %i.e., average treatment effect with the path $D\to M$ deactivated. 
which measures the average change in outcomes due to the treatment, while the mediator is kept at its level that would be realized when $D=d$. Since the mediator is held fixed at $M=M_d$, $NDE$ measures the effect that does not operate through $M$. To motivate $NDE_0$, say, imagine the status quo where everyone is untreated so that $Y^{pre}=Y_{0,M_0}$ for all. Now, suppose that the new policy requires that everyone gets treated, while a policy-maker possibly deactivated the path from $D$ to $M$ so that $M$ is kept unchanged. Then $Y^{post}=Y_{1,M_0}$ would be realized. $NDE_0=\E[Y^{post}-Y^{pre}]$ measures the effect of such policy change.
%footnote: cite?
Such definition of direct effect is called \emph{natural} direct effect due to \cite{pearl}, in contrast to the \emph{non-natural} or \emph{controlled} direct effect defined by $\E[Y_{1,m}-Y_{0,m}]$ for $m=0,1$. Unlike natural direct effects, controlled direct effects set the value of mediator to certain level, $m$.\footnote{That is, natural DE is important when we hypothesize any path-disabling intervention. This is in contrast to the variable-setting intervention where we set $M=m$ for everyone, as being hypothesized by the controlled DE (CDE). It is implied in the definition of CDE that a policy-maker can conceivably set $M$ to specific value for everyone regardless of $D$. Whether that is realistic or policy-relevant would depend on the context of studies.} While controlled direct effects can be of interest as well, we focus on the natural direct effects following the majority of the literature.

Similarly, \emph{natural} indirect effects are defined as follows:
\begin{eqnarray}\label{IE}
NIE_{\textcolor{red}{d}}=^{def}\E[Y_{\textcolor{red}{d},M_{1}}-Y_{\textcolor{red}{d},M_{0}}],\quad\for d=0,1
\end{eqnarray}
which measures the average change in outcomes when the value of mediator changes from the value that would be realized under the control state (i.e., $M=M_0$) to the value that would be realized under the treatment state (i.e., $M=M_1$) while the treatment is fixed at its reference level, $D=d$. %PS 소개할까..

%=============================================================================%
\subsection{Identification issues and sequential ignorability assumption}
%=============================================================================% 
% and also cross-world counterfactuals
Identification of natural direct and indirect effects is challenging. Even when the treatment is randomized, there is no guarantee that the mediator is exogenous. Since our aim is to understand implications of the mediator endogeneity, we maintain the assumption that the treatment is random in order to isolate the essence of the problem: 
\begin{assumption}[treatment exogeneity]\label{a0} For all $d,d'$ and $m$, let
$$\big\{ Y_{d,m},M_{d'}\big\} \indep D.$$
\end{assumption}
Hereafter, we use $\indep$ to denote a mean independence. 
This assumption is satisfied for instance when we have a randomized experiment where a treatment is randomly allocated across individuals. Recall that we implicitly condition on observable covariates. Thus the assumption also covers  observational studies where researchers can reasonably assume that the treatment is unconfounded after controlling for sets of covariates.

$M$ is endogenous if there exists a common factor simultaneously affecting both $M$ and $Y$. In order to identify the causal effect of $M$ on $Y$, it is necessary to control for all of these common factors.  The ``sequential ignorability" (SI) proposed by \cite{IKY} assumes that these common factors are all observables, and thus can be controlled. Specifically SI assumes that for all $d,d',m$, we have
$$M\indep Y_{d,m}|D=d' $$
after controlling for observable covariates. Under SI assumptions, natural direct and indirect effects are nonparametrically identified as shown in \cite{IKY}. 
 
%=============================================================================%
\section{Instrumental Variable Approach to Mediation Analysis}
%=============================================================================%
SI assumption requires that confounders affecting both $M$ and $Y$ are entirely observable. As admitted by \cite{IKY}, this assumption is rather strong --- The assumption cannot be proven and, in many cases, is difficult to justify. Typically,  an instrumental variable (IV) method is used when we want to identify the causal effect of an endogenous variable, where the endogeneity stems from possibly unobserved confounders. Thus recent papers have started to propose an alternative identification strategy based on IVs. %why binary?
We seek to understand the identification power of such IV in a mediation setting. To do so, we consider the case where we have an access to a binary instrument, $Z\in\{0,1\}$, satisfying the following assumption:

%---------------------------------------------------------------------------------------------------------------------------------------%
\begin{assumption}[binary instrument]\label{a1} There exists a  binary instrument, $Z\in\{0,1\}$, such that for all values of $d,d',m,z$, the following statements hold:
\begin{enumerate}
\item[(i)] randomization of the instrument:
$$ \big\{Y_{d,m},M_{d’},D\big\} \indep Z.$$
\item[(ii)] exclusion restriction: $Y_{d,m,z}=Y_{d,m}.$
\item[(iii)] relevance: $\Pr(M|Z=z) $ is a nontrivial function of $z$.
\end{enumerate}
\end{assumption}
%---------------------------------------------------------------------------------------------------------------------------------------%
 (i) requires that $Z$ is exogenous. Recall that we are implicitly conditioning on covariates. (ii) requires that $Z$ does not affect $Y$ directly. On the other hand (iii) requires that $M$ should be affected by $Z$. Taken together, (ii) and (iii) require that $Z$ affects $Y$ only through $M$.
 
With such $Z$ at hand, we now augment our potential mediator notation from $M_d$ to $M_{d,z}$. Observed mediator is $M=M_{D,Z}$. We also use the notation $M_{0}=M_{0,Z}$ and $M_1=M_{1,Z}$ so that 
\begin{eqnarray*}
&&\E[M_0]=\E[M_{0,Z}]=\E[M_{01}|Z=1]\Pr(Z=1)+\E[M_{00}|Z=0]\Pr(Z=0)\\
&&=\E[M_{01}]\Pr(Z=1)+\E[M_{00}]\Pr(Z=0)
\end{eqnarray*}
where the last equality follows from $M_{d,z}\indep Z$. 
%=============================================================================%
%\subsection{What does IV identify?}
%=============================================================================%
IV estimators for mediation are then defined by postulating the following linear regression model for $Y$ and $M$:
\begin{eqnarray*}
Y
&=&\begin{cases}
\beta_0+\beta_1M+u,& \for D=0\\
\alpha_0+\alpha_1M+u,& \for D=1
\end{cases}\\
\end{eqnarray*}
and
\begin{eqnarray*}
M
&=&\begin{cases}
\pi_0+\pi_1Z+v,& \for D=0\\
\tau_0+\tau_1Z+v,& \for D=1
\end{cases}\\
\end{eqnarray*}
Equivalently, we have the following system of linear equations for endogenous variables:
\begin{eqnarray}
&&Y=D(\alpha_0+\alpha_1M)+(1-D)(\beta_0+\beta_1M)+u,\label{lim1} \\
&&M=D(\pi_0+\pi_1Z)+(1-D)(\tau_0+\tau_1Z)+v\label{lim2} 
\end{eqnarray}
where it is assumed that
$$\E[u|D,Z]=\E[v|D,Z]=0. $$

Linear coefficients, $\theta=(\alpha,\beta,\pi,\tau)$, are estimated by running an IV regression of $Y$ on $M$ using $Z$ as an instrument, separately for each value of $D$. 
Let $\widehat\theta^{IV}$ denote the resulting IV estimator. Given $\widehat\theta^{IV}$, direct and indirect effects are estimated by
\begin{eqnarray}
&&\widehat{NIE}_0=\widehat{\beta}_1\big(\widehat{\pi}_0-\widehat{\tau}_0+(\widehat{\pi}_1-\widehat{\tau}_1)\widehat{\E}[Z]\big) \label{NIE0}\\
&&\widehat{NIE}_1=\widehat{\alpha}_1\big(\widehat{\pi}_0-\widehat{\tau}_0+(\widehat{\pi}_1-\widehat{\tau}_1)\widehat{\E}[Z]\big)\\
&&\widehat{NDE}_0=\widehat{\alpha}_0-\widehat{\beta}_0
+(\widehat{\alpha}_1-\widehat{\beta}_1)\cdot\big(\widehat{\tau}_0+\widehat{\tau}_1\widehat{\E}[Z] \big)\\
&&\widehat{NDE}_1=\widehat{\alpha}_0-\widehat{\beta}_0
+(\widehat{\alpha}_1-\widehat{\beta}_1)\cdot\big(\widehat{\pi}_0+\widehat{\pi}_1\widehat{\E}[Z] \big)
\end{eqnarray}
where $\widehat{\E}[Z]=\sum_{i=1}^n Z_i/n$ (see \cite{van}). Let $\theta^{IV}$ be the probability limit of $\widehat\theta^{IV}$. Probability limits of estimated mediation effects are given as follows:
\begin{eqnarray}
&&NIE_0^{IV}=\beta_1^{IV}\big(\pi_0^{IV}-\tau_0^{IV}+(\pi_1^{IV}-\tau_1^{IV})\E[Z]\big)\label{IE0}\\
&&NIE_1^{IV}=\alpha_1^{IV}\big(\pi_0^{IV}-\tau_0^{IV}+(\pi_1^{IV}-\tau_1^{IV})\E[Z]\big)\\
&&NDE_0^{IV}=\alpha_0^{IV}-\beta_0^{IV}+(\alpha_1^{IV}-\beta_1^{IV})\cdot\big(\tau_0^{IV}+\tau_1^{IV}\E[Z] \big)\\
&&NDE_1^{IV}=\alpha_0^{IV}-\beta_0^{IV}+(\alpha_1^{IV}-\beta_1^{IV})\cdot\big(\pi_0^{IV}+\pi_1^{IV}\E[Z] \big).
\end{eqnarray}
On the other hand, nonparametrically, $NDE_d$ and $NIE_d$, when augmented with $Z$, are defined as follows:
\begin{eqnarray}
&&NIE_0=\E[Y_{0,M_{1,Z}}-Y_{0,M_{0,Z}}]=\sum_{z\in\{0,1\}}\E[Y_{0,M_{1,z}}-Y_{0,M_{0,z}}]\Pr(Z=z)\\
&&NIE_1=\E[Y_{1,M_{1,Z}}-Y_{1,M_{0,Z}}]=\sum_{z\in\{0,1\}}\E[Y_{1,M_{1,z}}-Y_{1,M_{0,z}}]\Pr(Z=z)\\
&&NDE_0=\E[Y_{1,M_{0,Z}}-Y_{0,M_{0,Z}}]=\sum_{z\in\{0,1\}}\E[Y_{1,M_{0,z}}-Y_{0,M_{0,z}}]\Pr(Z=z)\\
&&NDE_1=\E[Y_{1,M_{1,Z}}-Y_{0,M_{1,Z}}]=\sum_{z\in\{0,1\}}\E[Y_{1,M_{1,z}}-Y_{0,M_{1,z}}]\Pr(Z=z)
\end{eqnarray}
Compare, for instance, $NIE_0^{IV}$ and $NIE_0$. We show that when the linear model (\ref{lim1} and \ref{lim2}) is correctly specified (i.e., the relationship is truly linear with an additive heterogeneity term), we have $NIE_d^{IV}=NIE_d$ and $NDE_d^{IV}=NDE_d$ for all $d=0,1$. However, while linearity assumption may be justified on the ground of discreteness of $M$ and $D$, the assumption of constant slopes is strong as it assumes that homogeneity of effects conditional on covariate which is likely to be violated when individuals select into $M$ based on their unobservable gains, a case of what \cite{heckmanm} calls ``essential heterogeneity". In such case, coefficients $(\alpha,\beta)$ are random even when we control for all observables, and worse, $(\alpha_i,\beta_i)$ may be correlated with $M_i$.

 Our aim is to understand how to interpret $NIE_d^{IV}$ and $NDE_d^{IV}$, and whether they  are informative about target parameters, $NDE_d$ and $NIE_d$, when the model is misspecified.

%=============================================================================%
\subsection{What does IV identify?}
%=============================================================================%

As a first step, the following lemma shows what $\theta^{IV}$ identifies:
%---------------------------------------------------------------------------------------------------------------------------------------%

\begin{lemma}[causal interpretation of $\theta^{IV}$]\label{l1}
\begin{eqnarray*}
\beta_1^{IV}=\frac{\E[Y_{0,M_{01}}-Y_{0,M_{00}}]}{\E[M_{01}-M_{00}]},\quad  \beta_0^{IV}=\E[Y_{0,M_0}]-\beta_1^{IV}\E[M_0],\\
\alpha_1^{IV}=\frac{\E[Y_{1,M_{11}}-Y_{1,M_{10}}]}{\E[M_{11}-M_{10}]},\quad \alpha_0^{IV}=\E[Y_{1,M_1}]-\alpha_1^{IV}\E[M_1]
\end{eqnarray*}
and
\begin{eqnarray*}
\pi_1^{IV}  =\E[M_{01}-M_{00}],\quad \pi_0^{IV}=\E[M_0]-\pi_1^{IV}\E[Z]\\
\tau_1^{IV}  =\E[M_{11}-M_{10}],\quad \tau_0^{IV}=\E[M_1]-\tau_1^{IV}\E[Z] 
\end{eqnarray*}
where
\begin{eqnarray*}
\beta_1^{IV}=\frac{\E[Y_{01}-Y_{00}|M_{01}>M_{00}]\Pr(M_{01}>M_{00})-\E[Y_{01}-Y_{00}|M_{01}<M_{00}]\Pr(M_{01}<M_{00})}{\Pr(M_{01}>M_{00})-\Pr(M_{01}<M_{00})}
\end{eqnarray*}
and
\begin{eqnarray*}
\alpha_1^{IV}=\frac{\E[Y_{11}-Y_{10}|M_{11}>M_{10}]\Pr(M_{11}>M_{10})-\E[Y_{11}-Y_{10}|M_{11}<M_{10}]\Pr(M_{11}<M_{10})}{\Pr(M_{11}>M_{10})-\Pr(M_{11}<M_{10})}
\end{eqnarray*}
See \ref{p1} for proof. 
\end{lemma}

%---------------------------------------------------------------------------------------------------------------------------------------%
%\subsubsection{Comparing $IE_0^{IV}$ with $IE_0$}
%---------------------------------------------------------------------------------------------------------------------------------------%
\newpage
Now, let us focus on the question of what $NIE_0^{IV}$ identifies ($NIE_1^{IV}$ case can be done in a symmetric way). From equation \ref{IE0}, we have
\begin{eqnarray}
NIE^{IV}_0&=&\beta_1^{IV}\underbrace{\big(\pi_0^{IV}-\tau_0^{IV}+(\pi_1^{IV}-\tau_1^{IV})\E[Z]\big)}_{=\E[M_1-M_0]}\\
&=&\beta_1^{IV}\sum_{z\in\{0,1\}}\E[M_{1z}-M_{0z}]\pi_z \label{t3}
\end{eqnarray}
while $NIE_0=\E[Y_{0,M_{1,Z}}-Y_{0,M_{0,Z}}]$ can be written as
\begin{eqnarray}
\sum_{z\in\{0,1\}}\Big[\E[Y_{01}-Y_{00}|M_{1z}>M_{0z}]\Pr(M_{1z}>M_{0z})-\E[Y_{01}-Y_{00}|M_{1z}<M_{0z}]\Pr(M_{1z}<M_{0z})\Big]\pi_z. \label{e19}
\end{eqnarray}
Suppose that the effect $Y_{01}-Y_{00}$ is constant across individuals. In such case, we have $\beta_1^{IV}=\E[Y_{01}-Y_{00}]$ and $NIE_{0}=NIE_0^{IV}$, so that IV estimator of $NIE_0$ identifies the true $NIE_0$. The result can be generalized to other (in)direct effects as well:
\begin{proposition}[IV estimand under constant effects]
Conditional on observables, if  $Y_{d',m'}-Y_{d,m}$ is constant across all individuals, then $NIE^{IV}_d=NIE_d$ and $NDE^{IV}_d=NDE_d$ for all $d=0,1$.
\end{proposition}

%=============================================================================
\subsection{Monotonicity Conditions}
In general, however, such constant effect assumption is hard to justify. It is violated when individuals with different values of potential mediator values, $\{M_{d,z}\}_{(d,z)\in\{0,1\}^{2}}$, experience systematically different (in)direct effects. Note that since $M_{d,z}$ is not observed for all possible values of $(d,z)$ (i.e., \emph{compliance type} is unknown), we cannot control for this. Once the (unobserved) effect heterogeneity is allowed, it is not clear whether and how $NIE_0^{IV}$ (eq. \ref{t3}) is comparable to $NIE_0$ (eq. \ref{e19}). 

%=============================================================================%
We argue that actually $NIE_0^{IV}$ does not have a causal interpretation under the heterogenous effect setting without making an additional  assumption on $M$, namely, monotonicity assumption.

Our first claim is that in fact, the target parameter $NIE_0$ itself has no casual interpretation under heterogeneous effects setting unless monotonicity of $M$ with respect to $D$ for given value of $Z$ is assumed. To see this, let us implicitly condition on $Z=z$. Note that
\begin{eqnarray*}
NIE_0&=^{def}&\E[Y_{0,M_1}-Y_{0,M_0}]\\
&=&\E[Y_{01}-Y_{00}|M_1>M_0]\Pr(M_1>M_0) -\E[Y_{01}-Y_{00}|M_1<M_0]\Pr(M_1<M_0)
\end{eqnarray*}
Thus, $NIE_0$ identifies a weighted difference of two effects: (i) $\E[Y_{01}-Y_{00}|M_1>M_0]$, an average $Y_{01}-Y_{00}$ for those with $(M_{0},M_1)=(0,1)$ and $\E[Y_{01}-Y_{00}|M_1<M_0]$, an average $Y_{01}-Y_{00}$ for those with $(M_{0},M_1)=(1,0)$. Since $NIE_0$ is a weighted difference of two different effects, we may have $NIE_0=0$ even when there is $Y_{0,M_1}-Y_{0,M_0}\not=0$ for everyone because two effects cancel each other. Thus we may wrongly conclude there is no mediation.

This problem occurs since some individuals change their mediator status from $0$ to $1$ when given the treatment, while some individuals just do the opposite and change their mediator value from $1$ to $0$. As a result, an overall impact of giving the treatment does not contain any information about causal effect of a treatment on any person. The problem here is analogous to the case that the Wald estimand does not identify any causal parameters when both compliers and defiers are coexisting. Similarly, the problem can be avoided when we assume that treatment affects $M$ in the same direction for everyone: %---------------------------------------------------------------------------------------------------------------------------------------%
\begin{assumption}[weak monotonicity of $M$ in $D$]\label{d-mono} for all $z\in\{0,1\}$,
$$\Pr(M_{1,z}\ge M_{0,z})=1 $$
\end{assumption}
%---------------------------------------------------------------------------------------------------------------------------------------%
This assumption requires that for given $z$, being treated only weakly increases the value of $M$. Under this assumption, equations \ref{t3} and \ref{e19} become
 \begin{eqnarray}\label{t5}
NIE_0&=&\sum_{z\in\{0,1\}}\E[Y_{01}-Y_{00}|M_{1z}>M_{0z}]P_{1z}\pi_z
\end{eqnarray}
and
\begin{eqnarray}\label{t6}
NIE^{IV}_0
&=&\frac{\E[Y_{0,M_{01}}-Y_{0,M_{00}}]}{\E[M_{01}-M_{00}]}\sum_{z\in\{0,1\}}P_{1z}\pi_z
\end{eqnarray}
where $\E[Y_{0,M_{01}}-Y_{0,M_{00}}]/\E[M_{01}-M_{00}]$ is
\begin{eqnarray*}
\frac{\E[Y_{01}-Y_{00}|M_{01}>M_{00}]\Pr(M_{01}>M_{00})-\E[Y_{01}-Y_{00}|M_{01}<M_{00}]\Pr(M_{01}<M_{00})}{\Pr(M_{01}>M_{00})-\Pr(M_{01}<M_{00})}
\end{eqnarray*}
For simplicity, let $Q_1=\Pr(M_{01}>M_{00})$ and $Q_2=\Pr(M_{01}<M_{00})$, so that equation \ref{t6} can be written as follows:
\begin{eqnarray}\label{t7}
NIE^{IV}_0
&=&\frac{\E[Y_{01}-Y_{00}|M_{01}>M_{00}]Q_1-\E[Y_{01}-Y_{00}|M_{01}<M_{00}]Q_2}{Q_1-Q_2}\sum_{z\in\{0,1\}}P_{1z}\pi_z
\end{eqnarray}
which, still, gives a non-convex combination of group-specific average effects.

By comparing equations \ref{t5} and \ref{t7}, we conclude that unless either $Q_1=0$ or $Q_2=0$, $NIE_0^{IV}$ is not informative about the $NIE_0$. We show this by using numerical example taken from \cite{IA2}: suppose $Q_1=2/3$ and $Q_2=1/3$ while $\E[Y_{01}-Y_{00}|M_{01}>M_{00}]=\alpha$ and $\E[Y_{01}-Y_{00}|M_{01}<M_{00}]=2\alpha$ with $\alpha>0$. Even when two subgroup-effects are positive, we would have $NIE_0^{IV}=0$. On the other hand, $NIE_0$ can take any sign. 
 
Again, we have the same problem as in \cite{IA2}: The effects for those mediator value is shifted from 0 to 1 when $Z$ is switched on can be cancelled out by the effects of those whose mediator value is shifted from one to zero. To avoid such problem, we again impose a moonotonicity assumption, this time for a given value of $D=d$:

%---------------------------------------------------------------------------------------------------------------------------------------%
\begin{assumption}[Weak monotonicity of $M$ in $Z$]\label{z-mono} for all $d\in\{0,1\}$,
$$\Pr(M_{d,1}\ge M_{d,0})=1 $$
\end{assumption}
%---------------------------------------------------------------------------------------------------------------------------------------%

%---------------------------------------------------------------------------------------------------------------------------------------%
Our final result shows that even when both assumptions \ref{d-mono} and \ref{z-mono} are satisfied, $NIE_0^{IV}$ identifies different quantity from $NIE_0$:  
\begin{eqnarray}
NIE_0=\sum_{z\in\{0,1\}} \E[Y_{01}-Y_{00}|M_{1z}>M_{0z}]\Pr(M_{1z}>M_{0z})\pi_z
\end{eqnarray}
while
\begin{eqnarray}
NIE_0^{IV}=\E[Y_{01}-Y_{00}|M_{01}>M_{00}]\sum_{z\in\{0,1\}} \Pr(M_{1z}>M_{0z})\pi_z
\end{eqnarray}
Here $NIE_0$ measures the overall effect for those who change their $M$ due to change in treatment value weighted over different values of $Z$. In contrast, $NIE^{IV}_0$ measures the one for those who change $M$ in response to $Z$ for a fixed $D=0$ world, multiplied by a constant $\sum_{z\in\{0,1\}}\Pr(M_{1z}>M_{0z})\pi_z$. 

 While we do not have non-convex weights problem anymore, so that both two have causal interpretations, it is not clear how $NIE_0^{IV}$ would be informative about the target parameter, $NIE_0$. Thus, it would be desirable to examine the degree of effect heterogeneity over different compliance group.  (Note that in the extreme case where there is no effect heterogeneity, these two are equivalent as expected.) Although we have focused on $NIE_0$ case, it follows easily that the same conclusion holds for $NIE_1$ as well as $NDE_d$. 

Our result thus implies that careful examination is needed in using IV methods when the target parameter is the form of natural (in)direct effects. While IV has a benefit of allowing unobserved confounders, its benefit comes with costs: either strong effect homogeneity or the monotonicity assumptions combined with concern of external validity is needed.   
 
%=============================================================================%

\section{Concluding Remarks} 
This paper investigates an identification of direct and indirect effects in a mediation setting using an instrumental variable. We have considered a simple case of a binary treatment, a binary mediator and a binary instrumental variable where there exists an unobserved confounder affecting both mediator and outcome. We have shown that the instrumental variable estimators based on linear models can identify natural direct and indirect effects when there is no unobserved heterogeneity in effects. Under the effect heterogeneity, we show that the instrumental variable estimators of natural direct and indirect effects do not deliver causally meaningful quantities without making certain sets of monotonicity assumptions restricting how the mediator responses to the treatment and the instrumental variable. We also show that even when these monotonicity assumptions are satisfied, the instrumental variable estimators do not necessarily correspond to natural (in)direct effects.

In conclusion, while IV methods have benefits of addressing unobserved confounders, caution would be needed. The comparative advantage of IV methods over traditional methods based on selection-on-observables would be lower as the degree of unobserved effect heterogeneity gets higher. Careful examination of the plausibility of homogeneity assumption along with sensitivity analysis with respect to unobserved effect heterogeneity would be fruitful.

%=============================================================================%
\newpage
\begin{appendix}
\section{Proof of Lemma \ref{l1}}\label{p1}  IV estimator is derived under the following conditions:
\begin{eqnarray}
\E[DZu]=0,\quad \&\quad  \E[Du]=0.
\end{eqnarray}
Since $u=Y-D(\alpha_0+\alpha_1M)-(1-D)(\beta_0+\beta_1M)$, above equations can equivalently be written as:
\begin{eqnarray}
\E[DZY]=\E[DZ(\alpha_0+\alpha_1M)],\quad  \&\quad
\E[DY]=\E[D(\alpha_0+\alpha_1M)].
\end{eqnarray}
Equivalently,
\begin{eqnarray}
\E[Y|D=1,Z=1]&=&\alpha_0+\alpha_1\E[M|D=1,Z=1],\\
\E[Y|D=1]&=&\alpha_0+\alpha_1\E[M|D=1]
\end{eqnarray}
which gives two equations with two unknowns with
\begin{eqnarray*}
&&\alpha_1=\frac{\E[Y|D=1,Z=1]-\E[Y|D=1]}{\E[M|D=1,Z=1]-\E[M|D=1]},\\
&&\alpha_0 = \E[Y|D=1]-\alpha_1\E[M|D=1]
\end{eqnarray*}
Using the fact that $\E[W|D=1]=\E[W|D=1,Z=1]\Pr(Z=1)+\E[W|D=1,Z=0]\Pr(Z=0)$ for any random variable $W$, $\alpha_1$ can be rewritten as follows:
\begin{eqnarray*}
&&\alpha_1=\frac{\E[Y|D=1,Z=1]-\E[Y|D=1,Z=0]}{\E[M|D=1,Z=1]-\E[M|D=1,Z=0]},\\
&&\alpha_0 = \E[Y|D=1]-\alpha_1\E[M|D=1]
\end{eqnarray*}
which can be expressed in terms of counterfactuals as follows:
\begin{eqnarray*}
\alpha_1&=&\frac{\E[Y_{1,M_{11}}-Y_{1,M_{10}}]}{\E[M_{11}-M_{10}]},\\
\alpha_0&=&\E[Y_{1,M_1}]-\alpha_1\E[M_1]
\end{eqnarray*}
Similarly, the expression for $(\beta_0,\beta_1)$ can be derived using the moment conditions: $\E[(1-D)Zu]=\E[(1-D)u]=0$:
\begin{eqnarray*}
\beta_1&=&\frac{\E[Y_{0,M_{01}}-Y_{0,M_{00}}]}{\E[M_{01}-M_{00}]},\\
\beta_0&=&\E[Y_{0,M_0}]-\beta_1\E[M_0].
\end{eqnarray*}

%=============================================================================%%=============================================================================%

\end{appendix}

%=============================================================================%

\newpage
\bibliographystyle{plainnat}
 %---> bibliog.bib contains bibtex information
\bibliography{IVmediation}

\begin{thebibliography}{13}
\providecommand{\natexlab}[1]{#1}
\providecommand{\url}[1]{\texttt{#1}}
\expandafter\ifx\csname urlstyle\endcsname\relax
  \providecommand{\doi}[1]{doi: #1}\else
  \providecommand{\doi}{doi: \begingroup \urlstyle{rm}\Url}\fi

\bibitem[Angrist and Imbens(1995)]{IA2}
Joshua~D. Angrist and Guido~W. Imbens.
\newblock Two-stage least squares estimation of average causal effects in
  models with variable treatment intensity.
\newblock \emph{Journal of the American Statistical Association}, 90\penalty0
  (430):\penalty0 431--442, 1995.

\bibitem[Chen et~al.(2019)Chen, Chen, and Liu]{chen}
Stacey~H. Chen, Yen-Chien Chen, and Jin-Tan Liu.
\newblock The impact of family composition on educational achievement.
\newblock \emph{Journal of Human Resources}, 54\penalty0 (1), 2019.

\bibitem[Dippel et~al.(2020)Dippel, Robert, Stephan, and Pinto]{dippel}
Christian Dippel, Gold Robert, Heblich Stephan, and Rodrigo Pinto.
\newblock Mediation analysis in iv settings with a single instrument.
\newblock working, 2020.

\bibitem[Dippel et~al.(2021)Dippel, Robert, Stephan, and Rodrigo]{dippel2}
Christian Dippel, Gold Robert, Heblich Stephan, and Pinto Rodrigo.
\newblock The effect of trade on workers and voters.
\newblock Economic Journal, accepted, 2021.

\bibitem[Fr{\"o}lich and Huber(2017)]{frolich}
Markus Fr{\"o}lich and Martin Huber.
\newblock Direct and indirect treatment effects--causal chains and mediation
  analysis with instrumental variables.
\newblock \emph{Journal of the Royal Statistical Society: Series B (Statistical
  Methodology)}, 79\penalty0 (5):\penalty0 1645--1666, 2017.

\bibitem[Heckman et~al.(2013)Heckman, Pinto, and Savelyev]{early}
James Heckman, Rodrigo Pinto, and Peter Savelyev.
\newblock Understanding the mechanisms through which an influential early
  childhood program boosted adult outcomes.
\newblock \emph{American Economic Review}, 2013.

\bibitem[Heckman(2001)]{heckmanm}
James~J. Heckman.
\newblock Micro data, heterogeneity, and the evaluation of public policy: Nobel
  lecture.
\newblock \emph{Journal of Political Economy}, 109\penalty0 (4):\penalty0
  673--748, 2001.

\bibitem[Imai et~al.(2010)Imai, Keele, and Yamamoto]{IKY}
Kosuke Imai, Luke Keele, and Teppei Yamamoto.
\newblock Identification, inference and sensitivity analysis for causal
  mediation effects.
\newblock \emph{Statistical Science}, 25\penalty0 (1):\penalty0 51--71, 2010.

\bibitem[Imai et~al.(2013)Imai, Tingley, and Yamamoto]{imai2013}
Kosuke Imai, Dustin Tingley, and Teppei Yamamoto.
\newblock Experimental designs for identifying causal mechanisms.
\newblock \emph{Journal of the Royal Statistical Society: Series A (Statistics
  in Society)}, 176\penalty0 (1):\penalty0 5--51, 2013.

\bibitem[Imbens and Angrist(1994)]{IA}
Guido~W. Imbens and Joshua~D. Angrist.
\newblock Identification and estimation of local average treatment effects.
\newblock \emph{Econometrica}, 62:\penalty0 467--475, 1994.

\bibitem[Mattei and Mealli(2011)]{mealli}
A.~Mattei and F.~Mealli.
\newblock Augmented designs to assess principal strata direct effects.
\newblock \emph{Journal of the Royal Statistical Society: Series B (Statistical
  Methodology)}, 2011.

\bibitem[Pearl(2001)]{pearl}
Judea Pearl.
\newblock Direct and indirect effects.
\newblock In \emph{Proceedings of the Seventeenth Conference on Uncertainty in
  Artificial Intelligence}, UAI'01, pages 411--420, San Francisco, CA, USA,
  2001. Morgan Kaufmann Publishers Inc.
\newblock ISBN 1558608001.

\bibitem[VanderWeele(2016)]{van}
Tyler~J. VanderWeele.
\newblock Mediation analysis: A practitioner's guide.
\newblock \emph{Annual Review of Public Health}, 37\penalty0 (1):\penalty0
  17--32, 2016.

\end{thebibliography}

\end{document}